\documentclass[10pt,letterpaper]{article}
\usepackage[top=0.85in,left=2.75in,footskip=0.75in]{geometry}

\usepackage{amsmath,amssymb}

\usepackage{changepage}

\usepackage{textcomp,marvosym}

\usepackage{cite}

\usepackage{nameref,hyperref}

\usepackage[right]{lineno}

\usepackage[nopatch=eqnum]{microtype}
\DisableLigatures[f]{encoding = *, family = * }

\usepackage[table]{xcolor}

\usepackage{array}

\usepackage{epigraph}

\usepackage{todonotes}
\reversemarginpar

\newcolumntype{+}{!{\vrule width 2pt}}

\newlength\savedwidth

\raggedright
\usepackage[aboveskip=1pt,labelfont=bf,labelsep=period,justification=raggedright,singlelinecheck=off]{caption}

\usepackage{lastpage,fancyhdr,graphicx}
\usepackage{epstopdf}
\fancyheadoffset[L]{2.25in}
\fancyfootoffset[L]{2.25in}
\begin{document}
\vspace*{0.2in}

\begin{flushleft}
{\Large
\textbf\newline{A CRISP approach to QSP: XAI enabling fit-for-purpose models} 
}
\newline
\\
Noah DeTal\textsuperscript{1\Yinyang},
Christian N. K. Anderson\textsuperscript{1\Yinyang},
Mark K. Transtrum\textsuperscript{1*\Yinyang}
\\
\bigskip
\textbf{1} Cross Stream Bioanalytics, Salt Lake City, UT, USA \\
\bigskip

\Yinyang These authors contributed equally to this work.
* mark@cross-stream.ai

Author Contributions: 
Conceptualization: NDD CNKA MKT
Data Curation: CNKA MKT
Formal Analysis: NDD CNKA MKT
Funding Acquisition: MKT
Investigation: NDD CNKA MKT
Methodology: NDD CNKA
Project Administration:	MKT
Resources: MKT
Software: NDD MKT
Supervision: MKT
Visualization:  NDD CNKA
Writing – Original Draft Preparation: NDD CNKA MKT
Writing – Review and Editing: NDD CNKA MKT

\end{flushleft}


\section*{Abstract}
Quantitative Systems Pharmacology (QSP) promises to accelerate drug development, enable personalized medicine, and improve the predictability of clinical outcomes.
Realizing this potential requires effectively managing the complexity of mathematical models representing biological systems.
Here, we present and validate a novel QSP workflow--CRISP (Contextualized Reduction for Identifiability and Scientific Precision)--that addresses a central challenge in QSP: the problem of complexity and over-parameterization, in which models contain irrelevant parameters that obscure interpretation and hinder predictive reliability.
The CRISP workflow begins with a literature-derived model, constructed to be comprehensive and unbiased by integrating prior mechanistic insights.
At the core of the workflow is the Manifold Boundary Approximation Method (MBAM), a reduction technique that simplifies models while preserving mechanistic structure and predictive fidelity.
By applying MBAM in a context-specific manner, CRISP links parsimonious models directly to predictions of interest, clarifying causal structure and enhancing interpretability.
The resulting models are computationally efficient and well-suited to key QSP tasks, including virtual population generation, experimental design, toxicology, and target discovery.
We demonstrate the utility of CRISP on case studies involving the coagulation cascade and SHIV infection, and identify promising directions for improving the efficacy of bNAb therapies for HIV.
Together, these results establish CRISP as a general-purpose QSP workflow for turning complex mechanistic models into tools for precise scientific reasoning to guide pharmacological and regulatory decision-making.

\section*{Author summary}
Quantitative Systems Pharmacology (QSP) uses mathematical models to understand complex biological systems and support drug development.
However, these models are often so complex that they limit precise scientific reasoning.
We introduce a new modeling workflow—CRISP (Contextualized Reduction for Identifiability and Scientific Precision)—designed to help scientists manage this complexity while improving the precision and reliability of QSP analyses.
Starting from comprehensive models built from published research, CRISP uses automated tools to simplify models based on specific scientific questions, preserving their ability to make accurate predictions.
This pipeline supports core QSP applications, including virtual population generation, experimental design, and target selection.
We demonstrate CRISP using case studies in blood coagulation and SHIV infection.


\section{Introduction}
\label{sec:intro}

Quantitative Systems Pharmacology (QSP) is an interdisciplinary field whose goals are to understand, predict, and control the behavior of complex biological systems\cite{sorger2011quantitative,musante2016quantitativ,zineh2019quantitativ}.
Detailed mathematical and computational models are at the heart of the QSP approach that, distinct from traditional approaches to drug development, incorporates information about mechanisms to elucidate the biochemical foundation of drug efficacy, toxicity, and resistance at the system level.
Such an approach incorporates diverse biological, pharmacological, and clinical data to create comprehensive models of drug action.
These models promise to make predictions under novel conditions, accelerate and reduce risk in drug development, and inform patient-specific treatment decisions in personalized medicine\cite{bai2020perspectiv,madabushi2022review}.
Because the biochemical processes underlying pharmacology are complex, many of the challenges faced by QSP modelers are direct consequences of the complexity inherent in the systems themselves, including computational cost, numerical stability and conditioning, data requirements, and unidentifiable parameters.
Consequently, there is no consensus workflow for building and validating QSP models\cite{helmlinger2019quantitativ,lemaire2023no,singh2023assessing,sokolov2025framework}, with many recent reviews emphasizing the need for techniques to address these issues within a QSP framework\cite{Sher2022,Noort2024,chung2023review}.
In this paper, we present and validate a novel QSP workflow--CRISP (Contextualized Reduction for Identifiability and Scientific Precision)--that addresses a central challenge in QSP: the problem of complexity and over-parameterization, in which models contain irrelevant parameters that obscure interpretation and hinder predictive reliability.
The pipeline is built on rigorous mathematical and deep philosophical foundations based on decades of careful research into the theory of predictive modeling using Information Geometry that endows researchers with an understanding of key mechanisms in their otherwise impenetrably complex systems.

\subsection{Existing Approaches to Managing Model Complexity}
Effective strategies to address complexity in systems theory are informed by the modeling objectives.
QSP models guide key decisions throughout the drug development process\cite{bai2020perspectiv}, often making predictions under novel conditions for which experimental data are not available and for which the relevance of diverse biological mechanisms is not known.
Consequently, complexity in QSP models tends to accumulate, often manifesting as a proliferation of unidentifiable parameters and plausible -- but indistinguishable --mechanistic alternatives.
It is widely recognized that effectively navigating complex models involves trade-offs between physical realism and parsimony\cite{hunt2018spectrum,Sher2022,lemaire2023no} with parameter identifiability being a central issue\cite{Noble2009,Noort2024}.
Parameter identifiability analysis methods include those based on Fisher Information\cite{eisenberg2014determinin,Noort2024} and profile likelihood\cite{Raue2009,eisenberg2014determinin}, while ensembles of unidentifiable parameters may be represented in virtual populations\cite{Schmidt2013,VPOP2016,Trogdon2025}.
A powerful approach involves rewriting the model parameters in terms of the identifiable combinations\cite{meshkat2014identifiab}.
Experimental design methods can determine maximally informative experiments within the limitations of the model\cite{casey2007optimal,apgar2010sloppy,chung2012experimental,white2016limitations}.

A range of model reduction strategies have been applied to QSP models, including lumping methods\cite{gulati2014scale,hasegawa2018automated}, time-scale separation\cite{peterson2012predicting,biswal2019structure}, 
conservation analysis\cite{vallabhajosyula2005conservation}, 
and control-theoretic techniques such as balanced truncation\cite{hahn2002improved,snowden2018model}.
In this work, we use the Manifold Boundary Approximation Method (MBAM) as our model reduction scheme\cite{transtrum2014model}.
It systematically removes parameter combinations that have minimal influence on model outputs, but retains both the predictive fidelity and mechanistic interpretability of the original model.
MBAM is unique among the above model reduction methods in that it does not assume a specific approximation structure \textit{a priori}; instead, it exploits the natural hierarchy of approximations intrinsic to the mathematical form of the model.
In this way, it recovers many of the approximations produced by other methods automatically as special cases\cite{transtrum2014model,transtrum2016bridging,pare2019model,transtrum2017measurement} based on information-theoretic criteria.
This flexibility has traditionally come at a large computational cost.
In this work, we leverage recent methodological advances that make MBAM applicable to large-scale QSP models.

Sensitivity analysis methods are also widely used in QSP to assess how model parameters influence system behavior.
Local sensitivity analysis (LSA), such as those based on the Fisher Information Matrix, are computationally efficient but can be misleading when sensitivity varies significantly across parameter space.
Global sensitivity analysis (GSA) techniques—such as Sobol’ indices~\cite{zhang_sobol_2015,Sobol}, Fourier-based methods (e/FAST)\cite{Homma1996}, and sampling-based approaches like Morris screening\cite{morris_factorial_1991}—address this limitation by estimating parameter influence over a broad range of values, but they become prohibitively computationally expensive in high-dimensional models.

Importantly, both LSA and GSA can obscure the mechanistic origins of the sensitivities they quantify.
A model may appear insensitive to certain parameters for reasons that range from biologically meaningful to mathematically trivial.
For example, if parameters are expressed in poorly scaled or unnatural units, both LSA and GSA may over- or under-estimate their importance.
In contrast, Information Geometry analysis, described below, provides a coordinate-invariant framework for quantifying parameter importance based on the intrinsic geometry of the model manifold.
This approach suggests a way forward that does more than suggest reductions: it certifies that the resulting reduced-order models retain predictive fidelity and reflect biologically meaningful simplifications, rather than artifacts of parameterization. 

\subsection{Information Geometry: A Theoretical Foundation for QSP}
The approach to managing complexity in QSP modeling presented here draws on ideas and techniques from the so-called sloppy model literature \cite{brown2003statistical,transtrum2015perspective,quinn2022information}.
The term sloppiness was first used to describe a phenomenon observed in systems biology models, where most parameters were poorly constrained by available data, yet the models nevertheless yielded reliable, precise predictions \cite{brown2003statistical,brown2004statistical}.
This counter intuitive behavior arises when predictions depend primarily on a few combinations of parameters, rendering most parameters effectively irrelevant to the outputs of interest.
Today, the term ``sloppiness'' remains as a misnomer, since mathematically rigorous methodologies based on Information Geometry now enable precise, mechanistic reasoning about the predictive capabilities of such models.
We refer to these collection of theories and techniques as ``Information Geometric analysis'' to emphasize the rigorous foundation on which they are built.

In our Information Geometry framework, parsimony is not a reflection of inherent simplicity in biological systems, but rather a pragmatic response to the fact that much of a model’s complexity is irrelevant for answering specific scientific questions.
This perspective has deep roots in statistical physics, where a rich tradition of theoretical work justifies simplified descriptions of complex systems \cite{anderson1972more,wilson1971renormalization,goldenfeld1999simple,laughlin2000cover,batterman2001devil,machta2013parameter,freeborn2023sloppy}.
Information geometry provides a rigorous foundation for identifying which aspects of a model matter most.
This foundation supports a suite of powerful techniques, including biologically grounded model reduction methods that yield parsimonious, fit-for-purpose models with rigorous performance guarantees \cite{transtrum2010why,transtrum2011geometry,transtrum2014model,transtrum2015perspective,transtrum2016bridging}.
These reduced models retain key mechanistic interpretability—such as the role of feedback loops—while offering lower computational cost, greater numerical stability, and clearer parameter identifiability and data requirements.
As such, Information Geometry provides an ideal foundation for precise reasoning about the links between biological mechanisms and system-level behavior in QSP applications.

In an Information Geometry analysis, a parameter's relevance is how accurately that parameter must be known for the model to make a precise prediction, as quantified by the Fisher Information Matrix (FIM).
Critically, parameter relevance is context-dependent, i.e. conditional on the specific predictions the model is used to make.
For any predictions of interest, we can partition a model's parameters into relevant and irrelevant combinations depending on whether they need to be finely tuned for the model to make precise predictions\footnote{This binary partition of parameters as either relevant or irrelevant is a conceptual simplification. In practice, parameters exist on a continuum of relevance depending on how finely they must be tuned to make predictions at a target precision.}.
This perspective is bolstered by biological intuition: phenomena of interest are often controlled by simple, understandable mechanisms of action embedded within the larger system, such as a key feedback loop or rate-limiting step.
As long as the parameter combinations associated with relevant mechanisms are accurately tuned, the remaining parameter combinations can vary appreciably without impacting the model's predictions.
This observation suggests that the irrelevant parameters and the processes they parameterize can be ignored in the appropriate context while maintaining the relevant mechanistic features of the model.

Of course, the challenges associated with complexity are not unique to QSP models; the search for parsimonious explanations is ubiquitous throughout science and engineering.
Historically, scientific modeling has been the art of separating relevant mechanistic details from irrelevant processes to arrive at the final model.
Such insights are often developed by domain-specific experts whose entire careers are devoted to studying a few specific systems.
Our most successful scientific theories back this intuition with mathematically rigorous abstractions.
For example, continuum limits allow us to ignore positions of individual atoms or molecules and instead work with spatially averaged concentrations.
In dynamical systems theory, singular perturbations abstract short-time dynamics and allow us to accurately model slower processes.
Many other examples could be given.
A major success of Information Geometry analysis has been the systematic demonstration that in these historically well-understood cases, Information Geometry consistently reproduces conclusions derived by expert intuition from diverse scientific domains.
However, it does so algorithmically by identifying the correct mathematical approximations that then guide intuition and new insights.
Successes include phase transitions in statistical field theories\cite{machta2013parameter}, continuum limits (i.e., spatial averaging of large numbers of molecules)\cite{machta2013parameter}, singularly perturbed dynamical systems\cite{chachra2012structural}, balanced truncation in control systems\cite{transtrum2017measurement,pare2019model}, and normal form theory in bifurcation analysis\cite{anderson2023sloppy}.
Accompanying these theoretical demonstrations are state-of-the-art technical tools for data fitting\cite{transtrum2010why,transtrum2012improvements}, Bayesian sampling\cite{girolami2011riemann,mattingly2018maximizing}, data visualization\cite{quinn2019visualizing,teoh2020visualizing}, and parameter reduction\cite{transtrum2014model,transtrum2016bridging,transtrum2017measurement}.
Information Geometry thus provides a suite of powerful computational tools built atop rigorous mathematical and philosophical foundations\cite{transtrum2015perspective,quinn2022information}.
These techniques collectively constitute a type of eXplainable Artificial Intelligence (XAI) that enable fit-for-purpose models in QSP and have the potential to support the QSP community in its goal of accelerating pharmaceutical development. 
By their nature, these tools should be applied judiciously depending on the specifics of the problem, and inform experts rather than replacing them. 

\subsection{Toward a CRISP Workflow}
A primary objective of the present work is to formalize a QSP workflow based on the principles and tools of Information Geometry.
As QSP emerges as a tool for model-informed drug development, the need for standardized workflows becomes increasingly evident\cite{bai2019translation,braakman2022evaluation}.
The dual issues of model granularity and parameter estimation are universally recognized as a central, open problem\cite{gadkar2016six,Ribba2017,helmlinger2019quantitativ,androulakis2022towards,ribba2023quantitative}.
In particular, balancing model complexity with parsimony can be challenging without a structured approach to scientific inquiry\cite{Ribba2017}.
Several authors have proposed structured, iterative workflows to address this gap.
Gadkar et al.\cite{gadkar2016six} introduced a six-stage framework emphasizing goal definition, biological scoping, model development, and iterative evaluation.
Helmlinger et al.\cite{helmlinger2019quantitativ} similarly highlighted the importance of cycles of refinement and validation grounded in empirical data.
Tools like the QSP Toolbox\cite{cheng2017qsp} have been developed to operationalize these frameworks, supporting tasks such as model calibration and virtual population generation.
Disease-specific applications, such as the workflow proposed by Go et al.\cite{go2024quantitativ} for atopic dermatitis trials, demonstrate how formalized QSP approaches can inform trial design and dosing strategies.
A standardized QSP workflow enables model credibility, regulatory alignment, and cross-disciplinary communication.

\begin{figure}
    \centering
    \includegraphics[width=\linewidth]{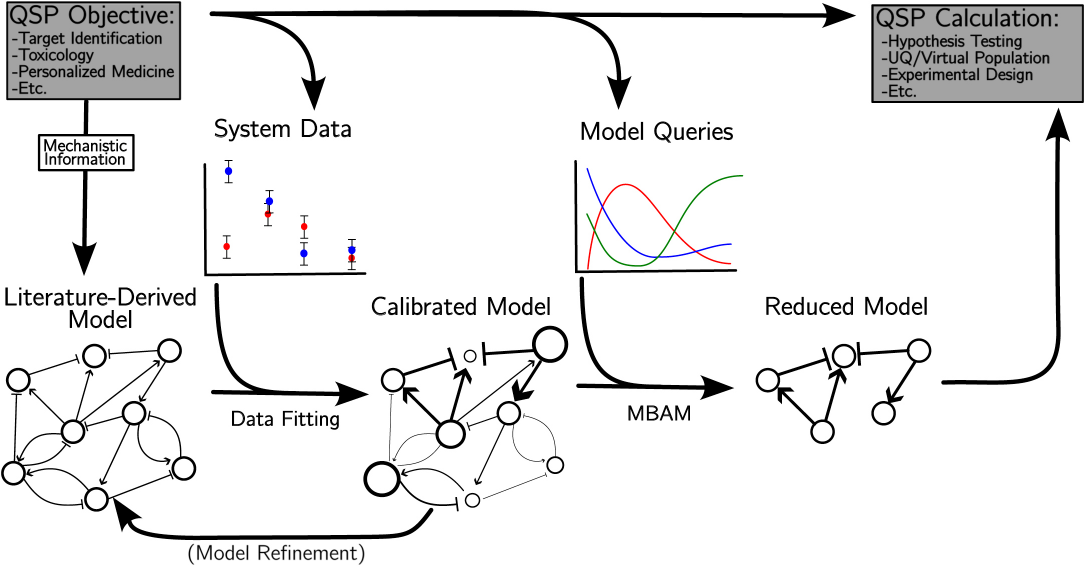}
    \caption{\textbf{Contextualized Reduction for Identifiability and Scientific Precision (CRISP) for QSP.}
      After building a detailed, literature-derived model, parameter values are calibrated by fitting to system data.
      Carefully selecting ``model queries'' based on the target application, a reduced-order model is built that removes irrelevant parameters and models the ``active manifold'' that efficiently propagates information from available data to predictions of interest.
      Calculations performed on the reduced model are mechanistically interpretable while having improved numerical conditioning, stability, and computational cost.
    }
    \label{fig1}
\end{figure}

We introduce a workflow leveraging Contextualized Reduction for Identifiability and Scientific Precision: CRISP.
The CRISP workflow is designed specifically to address issues related to model complexity and over-parameterization in the broader QSP framework.
As such, CRISP complements most other workflows, and key elements can be incorporated into other operational paradigms.
Key elements of CRISP procedure are illustrated in Fig~\ref{fig1} and briefly summarized here.  
A more detailed discussion is given in section~\ref{sec:SloppyWorkflow}.
Successful QSP flows from well-formulated questions; therefore, based on a specific QSP objective, e.g., target discovery, toxicology, etc., the modeling steps flow from this primary goal.
First, a detailed, mechanistic QSP model is constructed by uncritically combining mechanisms found in the relevant literature, and its (many) parameters are estimated from available data.
Next, specific modeling scenarios are designed, referred to as queries in our nomenclature (again, see section~\ref{sec:SloppyWorkflow} for longer discussion).
For example, one might be interested in predicting the system response under a novel drug regimen.
These queries are typically scenarios for which data is unavailable but for which one would like to use the model to make predictions.
Conceptually, the mathematical model is a machine that extracts information from the training data and propagates it forward to queries of interest.
However, the relationship between information in fitting data and future predictions is often obscured by the complexity of the literature-derived model, rendering many computational techniques ill-conditioned.
The next step, therefore, is to restrict the model to the parameters relevant to the queries.
We construct a reduced order model based on these queries using a Information Geometry algorithm called the Manifold Boundary Approximation Method (MBAM)\cite{transtrum2014model}.
By design, MBAM retains those parameters that are relevant to both the available data and the predictions of interest, clarifying the relationship between the two.
The reduced order model determines what information was learned from the data and how it propagates forward to inform the QSP objectives.
Critically, the model may not be identifiable from the data, but unidentifiable parameters in the reduced model reflect a fundamental limitation in the information available in the data.
Finally, subsequent QSP calculations--such as hypothesis testing, virtual population generation, or experimental design--are performed using the reduced model, and are informed by which modeling elements were retained by the reduction process.
In our experience, nearly all practical modeling tasks are improved, accelerated, and clarified by working with the parsimonious reduced model. 

The rest of this paper is organized as follows:
In section~\ref{sec:SloppyWorkflow} we present the proposed workflow in detail.
Since the key element of the workflow is the Manifold Boundary Approximation Method for model reduction, we next demonstrate this key step on a benchmark coagulation model in section~\ref{sec:coagulation}.
Finally, we demonstrate the workflow from beginning to end to study SHIV in section~\ref{sec:SHIV}.
Among other things, this workflow clearly identifies the most sensitive mechanisms for improving the efficacy of bNAb therapy.
This kind of mechanistic clarity is precisely what QSP aspires to deliver—enabling rational design, patient stratification, and regulatory confidence in model-informed drug development.

\section{Results}
\label{sec:Results}

\subsection{A CRISP QSP Workflow}
\label{sec:SloppyWorkflow}

A primary outcome of this work is the formalization of a QSP workflow based on principles and techniques of Information Geometry.
As summarized briefly in the introduction, the Contextualized Reduction for Identifiability and Scientific Precision (CRISP) workflow is illustrated graphically in Fig~\ref{fig1}.
Here we elaborate on this process before presenting the results of the validation studies.

In our approach, the quality of a mathematical model is always measured against its ability to facilitate answering specific scientific questions, illustrated by the box ``QSP Objectives'' in the upper left.
In QSP, the ultimate objectives are outcomes such as target discovery, toxicology, personalized medicine, and so forth.
The model realizes these high-level objectives through concrete calculations (box in upper right), such as testing a hypothesis by direct comparison against data, uncertainty quantification or virtual populations, designing experiments for further study, etc.
To develop a model fit for this purpose, we propose the workflow summarized in the flow diagram between these two boxes.

In the first step, a detailed model is constructed by gathering mechanistic structures from the available literature, resulting in what we term a literature-derived model.
In traditional practice, model building is as much an art as a science, relying heavily on the modeler’s physical and mathematical intuition to select structures that seem most relevant.
Yet in complex biological systems, intuition is slow to develop: assembling an initial model may take months, while true insight into system behavior often requires years or even decades.
Crucial modeling decisions—such as which mechanisms to include or omit—are shaped by expert judgment, but this judgment is inherently vulnerable to gaps in knowledge, disciplinary biases, and cognitive heuristics.
By contrast, the literature-derived model offers epistemic neutrality: it defers structural choices to the collective body of published research, ensuring that no single theoretical perspective dominates the model’s architecture.
In highly complex systems where causal pathways remain incomplete or ambiguous, this strategy guards against the universal human tendency to overemphasize the familiar while overlooking the novel or counterintuitive \cite{tversky_judgment_1974, gilovich_heuristics_2002}.
Moreover, by deliberately including all documented mechanisms upfront, our approach streamlines the modeling process: it reduces the time and expertise required for model construction, while anticipating the later application of automated, context-specific model reduction.
Rather than relying on human intuition to guess which mechanisms matter, the reduction step systematically identifies and preserves the structures essential to the phenomena of interest.
Literature-derived models can also be reused for later applications.
This shift standardizes the model building workflow, making it more automated, transparent, and objective, rendering it especially well-suited for scalable QSP applications.

Once realized in code, the literature-derived model can be thought of as a generic computational model since the parameters have not been calibrated to any specific data, and it can be computationally evaluated for a range of physically plausible parameter values corresponding to diverse model systems or even specific patients.
Therefore, the next step is to calibrate the parameters using experimental data.
In the flow diagram, the data to which the model is calibrated is labeled ``System data.''
This data is qualitatively different from the structural, literature-derived information used above.
The literature-derived model uses discrete knowledge about mechanisms to inform the mathematical and computational form of the model.
The experimental system data is continuous data used in regression analysis to calibrate the tunable parameters.
The result of this step is the ``calibrated model'' in Fig~\ref{fig1}.
Although not shown, one could have multiple calibrated models, for example, if data are taken for several distinct model systems, patients, etc.
In our SHIV study, we have multiple calibrated models for responsive, non-responsive, and untreated patients.

At this point, it is possible that the model cannot fit the data very well.
If that is the case, there are likely discrepancies between the mechanisms known in the literature and those of the real system.
If the fit is inadequate, the model could be further refined by adding additional terms to the model.
This was not necessary for any of the models used in this study; however, there are indications in our fits that models could be further improved in future work.
When a model fails to capture a target phenomenon, it does so transparently: not because it was skewed by human judgment, but because the requisite knowledge has not yet been formally articulated.
Thus, even the failure of a literature-derived model is highly informative, as it identifies an area where more fundamental science can be done.

Generally, we do not expect the parameters in the calibrated model to be tightly constrained by the available system data.
Results from Information Geometry demonstrate that QSP model parameters are generically unidentifiable.
However, a few combinations of parameters are well-constrained after fitting the data.
These are known as the stiff or relevant parameters, and these are the parameters that encode the information the model has taken from the system data.

Next, we further refine the purpose of the model by selecting which additional predictions we would like the model to make.
We refer to these as ``model queries.''
This verbiage is motivated by the concept that a model is a kind of knowledge repository to which we can pose questions.
For example, one might query a model by simulating a network of interacting proteins in response to a drug dose.
The predicted protein concentrations of interest at specific times are the model's response to the query.
The choice of model queries are motivated by what experimental data was available and what future predictions we would like the model to make.

The model queries are usually of two types: those for which experimental data is available and those for which it is not.
We represent these graphically by the different colored curves in Fig~\ref{fig1}.
The red and blue curves correspond to the model predictions for the red and blue experimental data points.
However, the green curve is meant to reflect a model prediction for a novel experimental condition for which experimental data are unavailable.

The purpose of the model is to propagate information from the experimentally constrained red and blue curves forward to the green curve.
The green curve will be well-constrained if it is only sensitive to those parameters that are constrained by the experimental data.
In practice, many parameters are unconstrained by experimental data, but they are also not relevant for the green curve.
Parameters that are both unconstrained by available data and not needed for other predictions are said to be irrelevant, and their presence complicates most QSP analyses.
Specifically, they make uncertainty propagation inherently ill-conditioned such that rigorous hypothesis testing, virtual population generation, experimental design, and any other task is computationally difficult at best and unreliable at worst.

To circumvent these challenges, we apply a parameter reduction algorithm known as the Manifold Boundary Approximation Method (MBAM)\cite{transtrum2014model}.
This algorithm carves reduced-order models from the literature-derived model, which include only parameters relevant to the given queries.
This reduced model is formally a special case of the full model; however, it is projected onto the relevant parameters that actively participate in the propagation of information from data to queries of interest.
To borrow the terminology of \cite{elton_applying_2021}, we apply Occam’s razor, but take care that it does not cut so deep the model loses its ability to ``reach'' behaviors for which we don't have experimental data~\cite{deutsch_beginning_2012}.
The query paradigm preserves parameters that are important not only for fitting available data, but also for generating reliable predictions, thus preserving relevant mechanisms even if they were unobserved experimentally.
We propose conducting the target QSP calculations using this reduced model to avoid the problem discussed above.
In the following sections, we demonstrate how these reduced models reveal the relevant control mechanisms underlying a behavior, are predictive under novel experimental conditions, and enable stable propagation of uncertainty while facilitating reasoning about mechanisms underlying different responses.

\subsection{Coagulation Modeling}
\label{sec:coagulation}

The coagulation cascade is a particularly well-studied system with characteristics that present typical challenges to QSP models, all of which are addressed by our workflow.
In this cascade, Factors VII (stable factor) and XII (Hageman factor) are activated by tissue damage, triggering two separate cascades which both activate Factor II (prothrombin), which in turn activates the clotting factor fibrin.
The underlying biology is well-characterized in the literature, with several previous studies having produced extensive literature-derived models.
It also exhibits many of the challenges and opportunities typical of QSP models.
The system spans multiple biological scales, from molecular (e.g., enzyme kinetics) to larger-scale physiological outcomes (e.g., clot formation).
It also exhibits strong nonlinearities and feedback mechanisms, which pose challenges for parameter identification, calibration, and sensitivity analysis.
On the experimental side, there is a wealth of \textit{in vitro}, \textit{in vivo}, and clinical data available for rigorous model validation and comparison studies.
Furthermore, many therapeutic agents, particularly anticoagulants, target the coagulation pathway, making the system relevant to realistic drug development scenarios.
In particular, models of the coagulation cascade are known to suffer from over-parameterization and have been used as validation studies for QSP workflows \cite{gulati2014scale,cheng2019clinical,chung2023review,ranc2023critical}.  

\begin{figure}[ht]
  \centering
  \includegraphics[]{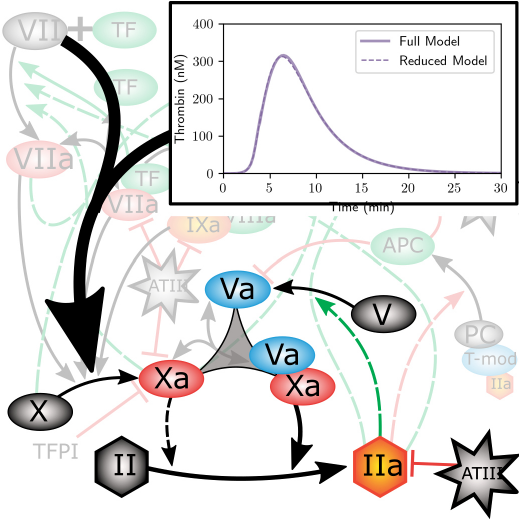}
  \caption{\label{fig2} 
  \textbf{Coagulation model.}
  Network diagram of the full (background) and reduced (foreground) coagulation cascade.
    The reduced model has 5 parameters and highlights the feed-forward loop, including the activation of Factor V by Factor IIa as the relevant mechanism.
    The thrombin response is adaptive, as shown in the inset for a typical profile.
  }
\end{figure}

In this study, we use the model proposed by Nayak et al.\cite{nayak2015using} as our base, literature-derived model whose primary reactions are summarized in the background of Fig~\ref{fig2}.
The model integrates the classical pathway elements into a detailed mechanistic model with an eye toward human-derived parameters for pharmacological applications. 
The model was validated using thrombin generation profiles (Factor IIa) and the system has been used in benchmarks for parameter identification, sensitivity analysis, and model reduction\cite{gadkar2016six,arumugam2017novel,hansen2019automated,neves‐zaph2024quantitativ,chen2025multistep}

\subsubsection{Minimal Model}
\label{sec:coagulation_minimal_model}

The parameter values reported by Nayak et al. give us a calibrated model; however, sensitivity analysis reveals that most of the parameters are unidentifiable from these data.
We wish to explore the ability of a reduced-order model to make predictions for QSP-specific queries.
The queries we chose are thrombin time series (Factor IIa) in response to various doses of Factor VIIa and Factor Xa added to both normal and Factor VII-deficient human plasma, for a total of 28 time series.
These are the same time series used in model calibration.
The thrombin response in each of these time series is adaptive\cite{ma2009defining,transtrum2016bridging}, i.e., after an initial response to the dosage, it returns to baseline levels after long times.
Our final coarse model based on these queries has five parameters, as is represented graphically in the foreground of Fig~\ref{fig2}, along with a typical thrombin profile for both the full and reduced models.

Notice how the reduced model dramatically highlights the key mechanism driving the adaptive behavior of the system.
In response to an initial dose, Factor X is activated, which then catalyzes the activation of Factor II.
This initial catalysis is relatively weak, but it subsequently triggers a feed-forward loop in which IIa activates Factor V, which then forms a complex with Xa.
The Va-Xa complex is a much stronger catalyst for activation of Factor II and is responsible for the bulk of the activation in each time series.
This causal narrative is consistent with how QSP scientists qualitatively understand the coagulation process.
Our pipeline algorithmically finds a minimal mathematical explanation from the literature-derived model that encapsulates the human intuition in a quantitative model with only five dynamical variables and four parameters.
The next section demonstrates the ability of the reduced model to answer key QSP questions.

\subsubsection{Predictive Power of the Minimal Model}
\label{sec:coagulation_predictions}

Recall that in our proposed workflow, the model queries are of two types: those for which experimental data is available and those for which it is not.
The rationale for this partition is that QSP models are often used to make out-of-distribution predictions for which data is not available, for example, system response under different dosages.
The mechanistic model effectively extracts information from the observed data through the calibrated parameter values.
It then propagates that information forward to generalize to unseen cases.
For this validation study, data is available for all the queries, which provides us with an opportunity to validate the informational relationships of different time series.
Here, we calibrate the reduced model to data for one of the 28 time series and then test its ability to predict key quantities of interest for the remaining 27 unobserved time series.

\begin{figure}[ht]
  \centering
  \includegraphics[width=0.98\textwidth]{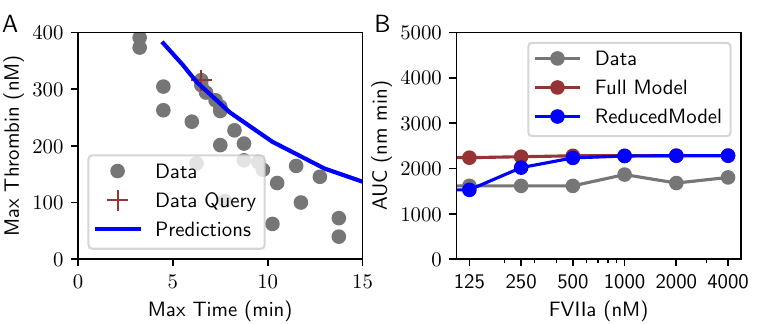}
  \caption{\label{fig3} 
  \textbf{Predictions of the reduced order coagulation model compared to data.}  
  The mechanism of action (incoherent feed-forward loop) guarantees that the max thrombin level is inversely proportional to the max time to achieve maximum thrombin activation (a).  
  Consequently, the area under the curve (b) is approximately constant over a wide range of control parameters.
  The model accurately predicts this trade-off, even when only one thrombin profile is used as a data query (a 94\% reduction in data demand).
  }
\end{figure}

Previous studies of the coagulation cascade used a few key statistical summaries of each thrombin profile: max time, max thrombin level, and area under the curve (AUC).
Having fit the reduced model to a single profile (i.e., the data query), we use it to predict these metrics for each of the remaining 27 time series and compare those predictions with those of the full model.
The results are summarized in Fig~\ref{fig3}.
Notice that because the reduced model preserves the mechanisms that are relevant for each of the queries, it accurately reflects the predictions of the full model.
Note the bias in the full model that consistently over-estimates the AUC is inherited by the reduced model.

It is noteworthy that the AUC (right panel of Fig~\ref{fig3}) is approximately constant for both the data and the models.
The reduced model clarifies the mechanistic reason for this invariance.
Indeed, the mechanism of action revealed by the reduced model is an incoherent feed-forward loop, one of two general mechanisms known to produce adaptive behavior\cite{ma2009defining}.
Previous studies showed that these simple mechanisms impose an inherent tradeoff between response time and response strength\cite{transtrum2016bridging} as is empirically observed for coagulation.
Understanding intrinsic constraints imposed by the mechanisms of actions are important to achieving QSP objectives.

The reduced-order models we present here offer several advantages over the full, literature-derived model for facilitating QSP studies.
First, they provide human-interpretable insights into the key mechanisms responsible for driving the phenomena of interest.
Beyond the biochemical simplifications evident from the reduced network structure, the fitted parameter values further highlight the relative importance of specific reactions in shaping the model’s predictions.
Second, the reduced models are computationally more efficient and numerically more stable than the full models.
In our study, direct simulation of the SBML-extracted full model revealed extreme sensitivity to numerical details: different integration algorithms produced wildly divergent solution profiles.
Only by evaluating multiple solvers at extremely high precision could we determine which solution accurately reflected the underlying mathematical equations.
By contrast, the reduced-order model achieved a two orders of magnitude reduction in computation time while simultaneously eliminating sensitivity to solver choice, enhancing numerical stability, and increasing overall confidence in model-based conclusions.
Notably, the process of addressing parameter identifiability through model reduction also resolves many of the broader challenges associated with model complexity.
We now illustrate these advantages in a case study on HIV infection dynamics, showcasing the full CRISP workflow from start to finish.

\subsection{SHIV-Modeling}
\label{sec:SHIV}

Having demonstrated how reduced models relate to the full calibrated model, and facilitate addressing QSP questions, we now demonstrate our end-to-end workflow for studying HIV infections.

\subsubsection{Literature-Derived and Calibrated Models}

\begin{figure}[ht]
  \centering
  \includegraphics[]{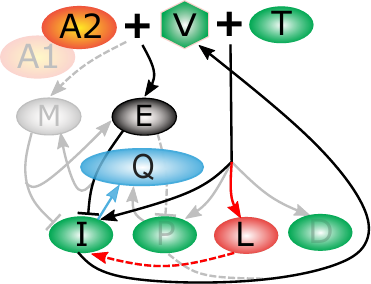}
  \caption{
    \textbf{SHIV Literature-derived model:} \textbf{V}iral load interacts with helper CD4+ \textbf{T}-cells, resulting in four classes of infected cells: actively \textbf{I}nfected, \textbf{L}atently, \textbf{P}ersistently, and \textbf{D}efectively infected cells. Such cells are camouflaged until the immune system is ``trained'' by broadly neutralizing antibodies (bNAbs) such as 3BNC117 (\textbf{A1}) and 10-1074 (\textbf{A2}). These drugs trigger the formation of long-lived \textbf{M}emory cells and far more immune \textbf{E}ffector cells than would potentially be generated naturally, which suppress actively and persistently infected cells with an efficiency depending on the level of immune exhaustion \textbf{Q}. Latently infected cells are able to reactivate and become actively infected cells. They are in red to denote that they are found in the nonresponder but not responder model; exhaustion is in blue, denoting its presence in the responder but not nonresponder model.
      }
  \label{fig4}  
\end{figure}

We first formulate a literature-derived model by aggregating known mechanisms from the published corpus.
As a starting point, we use the model of Desikan et al.~\cite{Desikan2020} to which we add several other mechanisms that have been suggested elsewhere in the literature\cite{Funk2001,Johnson2011,hutchison_modelling_2004,mclean_modelling_1994,Meyer2019}.
The goal of the literature-derived model is to incorporate every plausible mechanism present in the literature.
While the result is highly over-parameterized and surely includes irrelevant components, its construction is epistemically neutral; as explained in Section \ref{sec:SloppyWorkflow}, the model reflects the collective wisdom of the scientific corpus and does not preferentially emphasize some components over others.
Our literature-derived SHIV model has 11 species and 45 tunable parameters (see Methods) and is summarized by the network diagram in Fig~\ref{fig4}.

\begin{figure}
    \centering
    \includegraphics[]{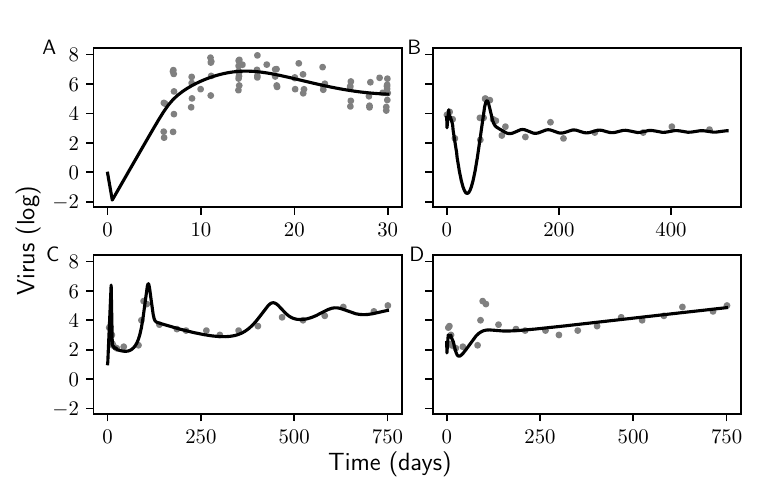}
    \caption{\textbf{Model fits} to the aggregate untreated data (A), responder DEWL data (B), and two fits to the nonresponder DF06 data showing a good fit (C) and an inferior but statistically significant fit that activates different mechanisms (D). 
    }
    \label{fig5}
\end{figure}

We calibrate the model by fitting it to data from the experiments performed by Nishimura et al.\cite{nishimura_early_2017} (see Methods).
We focus on fits to three data sets: viral load for 11 untreated monkeys, a responder labeled DEWL, and a nonresponder labeled DF06; all are shown in Fig~\ref{fig5}.

Note that previous studies used the terms ``controller'' and ``non-controller'' to indicate whether a specific monkey responded to treatment.
We instead use the terms ``responder'' and ``nonresponder'' to avoid confusion with the untreated comparison group, i.e., the statistical control.
Referring to all groups other than the control group using variants of the word ``control'' undermines the clarity of terminology.
Furthermore, ``elite controllers'' traditionally refers to individuals able to resist HIV \textit{without} intervention\cite{jacobs_cytokines_2017, berg_high_2021, hartana_long_2021, sugawara_learning_2022}.
For the purposes of this study, we do not wish to distinguish between innate resistance and resistance acquired through the bNAb intervention.

Although the model fits the viral load data reasonably well, there are naturally large uncertainties in the inferred parameters.
It is possible to find sub-optimal fits that are statistically significant that activate different mechanisms in the model.
This demonstrates that the relevant mechanism cannot always be uniquely identified by the data and that models often contain within them multiple competing hypotheses.
Accordingly, members of a population may exhibit different mechanistic responses to infection.
For the present study, we selected two parameter sets from the DF06 data that failed to respond to bNAb treatments for different reasons.
Our workflow identifies and clarifies these competing hypotheses.
These fits for two hypotheses are shown in the bottom row of Fig~\ref{fig5}.

The four parameter sets summarized by the fits in Fig~\ref{fig5} correspond to four distinct calibrated models in our workflow.
Each calibrated model becomes a starting point for model reduction in the next section.

\subsubsection{Reduced Models and Mechanistic Insights }
\label{sec:mechanisms}

We now seek reduced-order models for each of the four fits (calibrated models) in Fig~\ref{fig5}.
For model queries, we choose time series of change in T-cells, viral load, and bNAb concentrations $A_1$ and $A_2$ for no-, low-, medium-, and high-dose of bNAbs.
Notice that these queries include predictions for the data to which the model was fit (i.e., viral loads for untreated as well as for a medium bNAb dose).
We have extended the queries to include predictions for other dose regimes as well as for unobserved time-series of T-cells and $A_1$ and $A_2$ to better facilitate reasoning about the mechanistic relationship between viral load, T-cell, and bNAb level.
We are particularly motivated by the ability of the reduced-order models to give insight into key mechanisms of action underlying drug response that could guide human decision-makers in later stages of drug development.

Using these queries, we constructed reduced order models beginning from the parameter values for each of the four responses; the list of approximations and resulting equations for the resulting models are given in the methods section.
Just as for the coagulation model, we find the mechanistic insights revealed by these models to be useful for understanding the predictions of the model.
Unlike the coagulation cascade, however, we here have four reduced models, one for each calibrated model.
Both the similarities and differences among these models are insightful.

Many of the approximations are shared among all the reduced models.
Of particular interest is the reduced model that is generated by only those approximations common to all reductions\cite{petrie2022selecting}.
This model is known as the supremum because it is a kind of least upper bound---that is, it is the simplest MBAM-enabled model that contains all of the responses as special cases.
Because it is the simplest model that can explain all the data sets, the supremum model is particularly useful for reasoning about the relationships between the different models and the responses they describe.
For example, as the simplest model that can achieve all of the fits in Fig~\ref{fig5}, it is a natural model for designing interventions that could transform, for example, a nonresponder into a responder, as we do below in section~\ref{sec:targetid}.
Refer to the opaque parts of Fig~\ref{fig4} for the supremum model.

We now give a mechanistic, biological interpretation of the approximations in the supremum model.
While many approximations in the supremum have obvious biological meaning, others are more nuanced.
The removal of the defectively and persistently infected cells from the model indicates that these cell types play a negligible role in explaining infection dynamics on the time scale of 1-2 years.
This is unsurprising for defective cells, which are not expected to interact with disease dynamics once created.
On the other hand, the removal of the persistently infected compartment suggests surprisingly that the ``low profile'' maintained by some virally-transformed helper T-cells is a poor strategy for avoiding activated effector cells.

The removal of memory cells from the model suggests that they are not useful in preventing relapses or suppressing the ``second peak'' of viral load observed around day 60-80 in treated macaques (both responder and nonresponder alike).
It is possible they play a role in the rapid suppression of tertiary infection following anti-CD8 treatments, but we did not attempt to fit these data.
In other words, circulating levels of effector cells are sufficient to explain the suppression of secondary peaks, or the failure to do so in the case of nonresponders.

Although the pharmacokinetic predictions for both drugs are retained, the Hill function enhancing viral clearance is simplified to depend largely on $A_2$ for therapeutic benefit. 
Since the drugs are modeled to have symmetric effects, the presence of only a single drug is mathematically sufficient to predict the treated response. 
A more detailed model that distinguishes between the two drugs' mechanisms of action may provide more information about their synergy or competition. 

The approximations present in only responder or nonresponder models provide further insight into the mechanistic origins of these respective responses.
Understanding these differences opens the door for improved treatment protocols and ultimately personalized medicine.
The effector exhaustion level compartment $Q$ is present in responder, but not nonresponder, models. Sensitivity analysis of the responder model reveals that the viral reinfection peak is strongly influenced by the exhaustion dynamics. 
As a result, transient exhaustion is necessary to explain the observed behavior. 
The nonresponder, on the other hand, effectively saturates exhaustion throughout the duration of predictions, and so dynamic modeling can instead be replaced with constants.

The presence of latently infected cells in the nonresponder model—but not in the responder model—offers a mechanistic explanation for treatment failure.
The latent compartment serves as a reservoir for secondary reinfection, ultimately overwhelming the host in nonresponders.
In contrast, responders appear to either prevent the formation of latent cells or eliminate them efficiently upon activation into the actively infected compartment.
This finding aligns with prior work by Desikan et al.~\cite{Desikan2020}, who showed through extensive numerical experiments on manually curated models that latent cell dynamics were not necessary to explain responder behavior.
Our workflow reached the same conclusion automatically while simultaneously clarifying the critical role of latent cells in nonresponders.

These results suggest that targeting latent cells could be a viable strategy for converting nonresponders into responders.
Several therapeutic agents are known to activate latent cells—thereby rendering them vulnerable to treatment—including Bryostatin 1~\cite{mehla_bryostatin_2010}, synthetic analogues~\cite{marsden_vivo_2017}, and the recently discovered tigilanol EBC-46~\cite{gentry_synthesis_2025}.
Identifying such actionable mechanisms is a central goal of QSP, and the next section illustrates how our workflow systematically uncovers therapeutic targets across a wide range of models.
While latent cells represent a plausible target, our analysis reveals they may be suboptimal—other more sensitive mechanisms may elicit a stronger response, highlighting the value of an unbiased literature-derived model to prioritize interventions.

\subsubsection{Drug target identification}
\label{sec:targetid}

The insight from the previous section suggests potential mechanistic origins differentiating responders and nonresponders.
A natural extension of this analysis is target discovery: are there interventions that could transform nonresponders into responders?
The supremum is the natural model on which to conduct this analysis since it is the simplest model exhibiting both clinical outcomes.
We formulate a sparse optimization procedure (see methods) that identifies a minimal set of parameters that could be weakly perturbed to induce a controlled response in a nonresponder model.
These parameters and their associated mechanisms therefore serve as potential targets.

\begin{figure}
    \centering
    \includegraphics[]{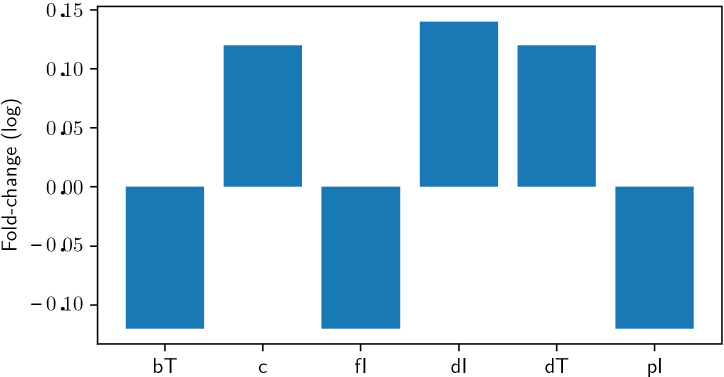}
    \caption{\textbf{Drug target parameters} for inducing a clinical response in a nonresponder. 
    The bars show the fold-change required for each individual parameter to induce response, with no change exceeding 15\% of the nominal value.}
    \label{fig6}
\end{figure}

Fig~\ref{fig6} displays the full list of individual target parameters and the log-change necessary to induce a clinically significant change in response from each one. 
Post hoc, the targets identified here are all mechanistically reasonable, e.g., increasing viral clearance $c$ intuitively improves the clinical outcome.
However, what is not intuitive \textit{a priori} is that the patient response would be sufficiently sensitive to these parameters to make them viable intervention targets. 
In the previous section, we discussed how targeting latent cells is a potential treatment strategy. 
However, our model shows that it requires a very large increase to the latent reactivation parameter $a$ to achieve clinical improvement, making it an unreasonable target for clinical therapy.
Other potential therapies that are mechanistically plausible from the full network similarly require extreme perturbations to achieve the desired effect.
For example, a 10,000-fold decrease in initial viral load $V_0$, has a negligible effect on the clinical outcome despite also being an intuitive target parameter.
Indeed, the full model includes several other potential targets that would be mechanistically reasonable but end up being irrelevant (i.e., too insensitive) for this clinical outcome. 
By performing our optimization procedure on the reduced model, we efficiently identify these sensitive, promising targets that achieve a desired response with changes less than 15\%.

\subsubsection{Uncertainty Quantification and Virtual Populations}
\label{sec:uq_oed}

A major advantage of QSP modeling is the ability to perform rigorous uncertainty quantification (UQ) to support regulatory decision-making.
UQ is particularly valuable for evaluating benefit-risk profiles, justifying dose selection, and informing the design of confirmatory studies.
However, when models are structurally unidentifiable, UQ analyses may become numerically unstable and computationally infeasible, limiting their utility.
To overcome these limitations, we apply our model reduction workflow to generate a lower-dimensional, identifiable model amenable to Bayesian UQ and virtual population generation.
This approach facilitates transparent characterization of variability in predicted outcomes and supports regulatory expectations for model credibility, including sensitivity to assumptions and relevance to the intended context of use.

We ask: given a representative response to untreated viral infection, can we characterize and quantify the range of responses to therapeutic intervention across a broader virtual population--\emph{prior} to conducting a clinical trial?
As a starting point, we use the parameter set calibrated to aggregated untreated data (Figure~\ref{fig4}a) to represent the nominal disease trajectory.
The model outputs of interest are viral load and T-cell time series under both untreated and bNAb-treated conditions.
The untreated response serves as the reference data query, while the treated trajectory represents the target prediction.
Our objective is to quantify uncertainty in these treatment outcomes and generate a virtual population reflecting plausible inter-individual variability in treatment response, thereby providing model-informed evidence to support benefit-risk assessments.
This approach aligns with the regulatory emphasis on establishing a clear context of use for quantitative systems pharmacology models to inform drug development decisions and to support regulatory submissions by characterizing clinically relevant variability and informing trial design.

MBAM gives a 16-parameter reduced model for these queries (Eq.~\eqref{eq:SHIVUTUQreduced} in the methods section).
Since these queries only include viral load predictions, a subset of those used in section~\ref{sec:mechanisms}, the model is even simpler than before.
Inspecting the resulting equations (see methods), six of the remaining parameters characterize the bNAb dynamics and therefore are trivially structurally unidentifiable from the untreated profile.
For now, we hold these at their nominal values and consider the uncertainty in the remaining ten parameters.
We then used this reduced model to generate a virtual population using Bayesian posterior sampling and categorized each virtual patient by its response both with and without treatment (see methods for more details). 

Our response categorization builds on prior QSP analyses\cite{Baral2019,Desikan2020} that showed that effective treatment leverages bistable dynamics in the model. 
We define successful treatment as shifting the equilibrium from the higher to the lower viral load steady state.
Accordingly, we divide the virtual population into two broad groups: those whose parameter sets produce monostable dynamics (a single stable equilibrium) and those that exhibit bistability.
The majority of virtual patients fall into the monostable category, meaning their dynamics converge to a single equilibrium that is not altered by treatment. 
While these patients may show a transient response, their long-term behavior remains unchanged, leaving little opportunity for long-term therapeutic benefit.

\begin{figure}
    \centering
    \includegraphics[width=\textwidth]{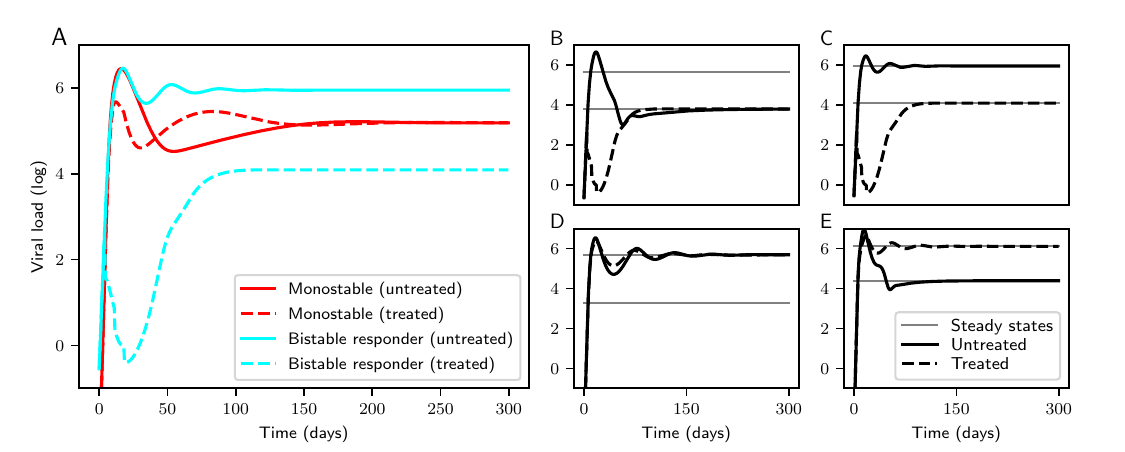}
    \caption{\textbf{Typical responses for each subpopulation}.
    (A) The virtual population is first partitioned into patients whose dynamics exhibit monostability and those that exhibit bistability.
    Among the bistable patients, four subpopulations are identified as either (B) immune, i.e., low viral count both with and without treatment, (C) responders, i.e., high viral count without treatment, low viral count with treatment, (D) non responders, i.e., high viral count with and without treatment, and (E) adverse, i.e., low viral count without treatment, high viral count with treatment.}
    \label{fig7}
\end{figure}

Bistable patients, in contrast, have two stable steady states: one corresponding to high viral load (progressive disease) and the other to low viral load (controlled disease).
We further subdivide this group into four categories based on their response to treatment.
Responder and nonresponder patients both converge to the high viral load state in the absence of treatment, but diverge under therapy: responders transition to the low viral load state with treatment, while nonresponders do not.
Immune patients naturally reach the low viral load state without intervention, and treatment has no effect on their outcome.
In contrast, adverse responders begin in the low viral load state without treatment, but therapy shifts their dynamics toward the high viral load state.
Figure~\ref{fig7} illustrates representative dynamics for each of these five subpopulations.

Table~\ref{Table1} summarizes the distribution of response categories within our 7,200-member virtual population.
Notably, all five subpopulations are represented.
However, only about 7\% of patients are classified as responders who benefit from bNAb intervention. 
In contrast, fewer than 0.2\% exhibit an adverse response, transitioning to higher viral load with treatment.
The predicted steady state for most patients is unaffected by therapy, indicating that either they were monostable, i.e., having only a single steady state, or that the treatment did not perturb the dynamics sufficiently toward the other steady state, corresponding to either non-responders or immune.

\begin{table}[h]
    \centering
    \begin{tabular}{|c|c|}
        \hline
        Monostable & Bistable \\ 
        \hline
        5683 & 1517 \\
        \hline
         & \begin{tabular}{@{}c|c|c|c@{}}
         Responder & Nonresponder & Immune & Adverse \\
         \hline
         486 & 576 & 445 & 10 \\
         \end{tabular} \\
         \hline
    \end{tabular}
    \caption{Categorization of the 7200-member virtual population generated from Bayesian posterior sampling. 
    Patients are first separated by monostable or bistable dynamics.
    Bistable patients are further separated by response both with and without bNAb therapy.}
    \label{Table1}
\end{table}

The statistics of the virtual population, such as the relative size of each subpopulation, are determined by the prior distribution used in the Bayesian sampling process.
For this study, we employed a flat prior over the logarithm of the model parameters, centered on the nominal values obtained by fitting to aggregated untreated data.
The prior spans 16 orders of magnitude in log-space, corresponding to a fold-change of nearly 10 million.
For this study, we have selected a prior that is intentionally broad to sample the full range of possible responses.
Roughly speaking, the relative distribution of the subpopulations reflects the volume of parameter space corresponding to each category.
As such, it does not necessarily reflect the relative distribution of responses in real patients whose parameters are unlikely to be uniformly distributed, nor so broad.
Importantly, the Bayesian framework applied to the reduced model allows us to sample efficiently from the posterior for any given prior, ensuring that all assumptions are explicitly defined at the outset.
In some applications, prior information about the natural distribution of parameters or responses could be used to construct different virtual populations.
This transparency represents a significant step forward in the standardization and reproducibility of virtual population generation.

\begin{figure}
    \centering
    \includegraphics[width=\textwidth]{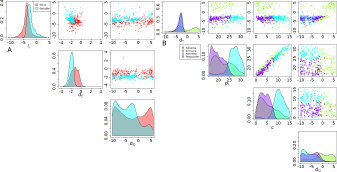}
    \caption{\textbf{Subpopulations and various parameter values}.
    (A) A random forest classifier identifies the parameters $\xi$, $d_E$, and $d_Q$ as the strongest predictor for whether a patient will be monostable or bistable.
    (B) The parameters $d_T$, $p_I$, $c$, and $d_Q$ are most useful for predicting whether a bistable patient will be a responder, non-responder, immune, or adverse.}
    \label{fig8}
\end{figure}

Because the virtual population is derived from a minimal model, we can directly identify which parameters determine the category to which a particular virtual patient belongs.
This, in turn, enables mechanistic reasoning about the origins of divergent responses and can inform strategies for monitoring safety during clinical trials or treatment stratification.
Using a random forest classifier, we find that the parameters $\xi$, $d_E$, and $d_Q$ are most strongly associated with the separation between monostable and bistable patients, as shown in Figure~\ref{fig8}a.
This suggests that the bifurcation separatrix is most sensitive to variation along these parameter directions. 
A 10\% cross-validation of this model could accurately predict mono- vs bistability for 94.5\% of out-of-training-set simulations ($\kappa=0.83$). 

Among the bistable patients, the parameters $d_T$, $p_I$, $c$ and $d_Q$ were most predictive of subpopulation membership (Figure~\ref{fig8}b).
Cross-validation of a random forest classifier trained on these parameters yielded 89\% accuracy in predicting bistable subtyptes ($\kappa=0.83$).
Due to the extremely low prevalence of adverse responders in our virtual population (0.13\%), there were insufficient data to robustly identify the specific parameter combinations driving adverse outcomes.
Nevertheless, qualitative analysis reveals that the basin of attraction associated with immune control is consistently narrow in these individuals.
This narrow basin explains both the rarity of adverse responders and the susceptibility of their trajectories to be perturbed into the progressive disease state.
These findings suggest that transient therapeutic perturbations—such as CART therapy or transient effector cell suppression—could shift system dynamics across the separatrix and redirect trajectories toward the healthier equilibrium.

From a clinical and regulatory perspective, the virtual population also supports safety evaluation and early biomarker identification.
Across all responder patients, CD4+ T-cell counts remained stable at over 90\% of uninfected levels over the full 1,000-day simulation period.
In contrast, other subpopulations—particularly non-responders and adverse responders—exhibited rapid T-cell decline to 10–80\% of baseline levels, often within the first 30 days post-infection.
These model-based predictions highlight the utility of early T-cell monitoring as a prognostic biomarker.
Specifically, CD4+ T-cell counts during the initial month post-infection may serve as a reliable surrogate endpoint for evaluating long-term therapeutic efficacy of bNAb-based interventions.

\section{Discussion}

Quantitative Systems Pharmacology (QSP) is a rapidly growing field with significant untapped potential to transform pharmacology\cite{braakman2022evaluation}.
By leveraging detailed mechanistic insights into biological systems, QSP holds promise for accelerating drug development, enabling personalized medicine, and shortening the path from discovery to deployment.
Realizing these goals, however, requires state-of-the-art analytical and computational tools that can extract maximum insight from both data and models.
In this work, we introduced and validated the Contextualized Reduction for Identifiability and Scientific Precision (CRISP) workflow for QSP that combines literature-derived models with a principled model reduction strategy based on the Manifold Boundary Approximation Method (MBAM).
Derbalah et al.\cite{derbalah2020framework} list eight characteristics of an ``ideal'' QSP model reduction system: system agnostic, automatable, mechanistically relevant, accurate, computationally low-cost, extrapolates, requires no [additional] experimental data, and estimates parameters.
Traditional MBAM has been limited in QSP deployment by computational expense; however, with advances in computational efficiency, MBAM now fulfills all eight criteria while also offering explainability at all levels of reduction.
Our workflow integrates these capabilities into a coherent process, providing a practical and explainable approach to modeling in mainstream QSP applications.

Several key features make our approach particularly well-suited for QSP applications.
First, the workflow is grounded in the well-established principles of Information Geometry, which provide a rigorous theoretical foundation and leverage cutting-edge numerical techniques.
Second, our use of a literature-derived model draws on the collective knowledge of the scientific community, helping to reduce bias in model structure and expand coverage of relevant mechanisms.
It simultaneously provides a tool to combat Kahneman's ``what you see is all there is'' bias\cite{tversky_judgment_1974, gilovich_heuristics_2002}, and a method for detecting knowledge gaps in the primary literature.
Finally, by generating context-specific reduced-order models tailored to target scientific or clinical queries, our approach ensures that predictive accuracy is retained while offering QSP scientists deeper insight into how the system functions.

The algorithmic nature of CRISP makes the modeling process largely automatic, yielding a form of eXplainable AI (XAI) for generating fit-for-purpose models.
This answers the call from the community to standardized procedures for deciding on model granularity and parameter estimation\cite{Ribba2017,gadkar2016six,bai2019translation,helmlinger2019quantitativ}.
As demonstrated, the CRISP workflow improves computational efficiency and numerical stability while providing modelers high-level insights into system behavior.
By automating key steps in model development, we hope to free QSP scientists from routine technical burdens, allowing them to focus their creative energies on scientific interpretation, hypothesis generation, and experimental design—ultimately helping to realize the full potential of the QSP paradigm.

Taken together, the CRISP workflow offers a practical and principled strategy for advancing QSP modeling in complex biological systems.
By integrating literature-derived models, Information Geometry, and query-driven model reduction, we provide a framework that is both computationally efficient and scientifically transparent.
Importantly, the modular and explainable nature of the workflow makes it well-suited for integration into clinical decision-making pipelines and regulatory review processes, where interpretability, traceability, and predictive reliability are paramount.
As QSP continues to expand its role in translational science, tools like this will be essential for scaling insight alongside complexity—and for delivering on the promise of mechanistic modeling in real-world pharmacological applications.

\section{Methods}

\subsection{Manifold Boundary Approximation Method}

The key technology that enables our workflow is the Manifold Boundary Approximation Method (MBAM).
In this approach, the query predictions of the literature-derived model for different parameter values are treated as a Riemannian manifold with the Fisher Information as the metric.
From an initial parameter vector, $\theta_0$, the FIM is calculated and an eigenvalue analysis is conducted.
Geodesics are calculated on the model manifold as described in reference\cite{transtrum2014model} originating from $\theta_0$ in the least sensitive parameter direction.
Calculating the Fisher Information matrix and then solving the geodesic equation involve calculating the sensitivity of the predictions with respect to the parameters.
These calculations are done using automatic differentiation within our in-house Julia toolbox.

At each iteration, the parameter values are monitored along the geodesic until a limiting approximation is detected.
Typical limiting approximations include things like equilibrium approximations (e.g., when the reaction rates are very large), turning of reactions (e.g., when rates get very small), and insaturable approximations (e.g., when Michaelis-Menten constants becomes very large).
The approximation is applied analytically to the equations to derive the reduced order model.

MBAM is described in detail with several systems biology examples in references\cite{transtrum2014model, transtrum2016bridging}.

\subsection{Coagulation}
\label{sec:coagulationmethods}

We use the model presented in reference \cite {nayak2015using} as the literature-derived model and the reported parameter values for the calibrated model.
We use as queries the 28 time series in Figure 2, panels B, G, L, and Q.
The equations for the reduced-order model are:
\begin{equation}
\label{eq:CoagulationModel}
\begin{split}
\dot{\widetilde{X}}_a &= \theta_1 \cdot VII_a - \widetilde{V}_a \cdot \widetilde{X}_a \\
\dot{II}_a & = \theta_2 \cdot II \cdot \widetilde{X_a V_a} + II \cdot \widetilde{X}_a - \theta_3 \cdot II_a \\
\dot{II} & = -\theta_2 \cdot II \cdot \widetilde{X_a V_a} - II \cdot \widetilde{X}_a \\
\dot{\widetilde{X_a V_a}} & = \widetilde{X}_a \cdot \widetilde{V}_a \\
\dot{\widetilde{V}}_a & = \theta_4 \cdot II_a
\end{split}
\end{equation}
where the initial conditions are zero except for $II$, which is a tunable parameter, and $\widetilde{X}_a$, which is set by the experimental protocol.  
$VII_a$ is also set by the experimental protocol.
Tildes represent "renormalized" states that are combinations of the full model states and unidentifiable parameters:
\begin{equation}
    \label{eq:CoagulationDefinitions}
    \begin{split}
        \widetilde{X}_a & = X_a \cdot k_{16} \\
        \widetilde{V}_a & = V_a \cdot k_{28} \\
        \widetilde{X_a V_a} & = X_a \cdot V_a \cdot k_{16},
    \end{split}
\end{equation}
where the $k$'s refer to the parameter labels in the original model.
The reduced parameters similarly have expressions in terms of the original parameters:
\begin{equation}
    \label{eq:CoagulationParameters}
    \begin{split}
        \theta_1 & = k_{63} \cdot X_0 \cdot k_{16} / K_{49} \\
        \theta_2 & = k_{31} / ( k_{16} \cdot K_{18} ) \\
        \theta_3 & = k_{41} \cdot ATIII_0 \\
        \theta_4 & = k_{26} \cdot k_{28} \cdot V_0
    \end{split}
\end{equation}

\subsection{SHIV}

\subsubsection{Literature-Derived Model}
\label{sec:SHIVmethods_litmodel}

Our literature-derived model combines multiple mechanisms from the literature and is given by Equation~\ref{eq:SHIVlitmodel}.
It models 11 individual species and has 45 tuneable parameters.
When not specified, the species are taken as zero at $t=0$.
\begin{equation} 
  \label{eq:SHIVlitmodel}
  \begin{split}
    \dot{T}&=b_T-d_T T-\left(f_I+f_L+f_P+f_D\right) T V \\
    \dot{V}&=p_I I+p_P P-(c+A) V \\
    \dot{I}&=f_I T V-d_I I+a L-m_I E I \\
    \dot{L}&=f_L T V-d_L L-a L \\
    \dot{P}&=f_P T V-d_P P-m_P E P \\
    \dot{D}&=f_D T V-d_D D \\ 
    \dot{E}&=\lambda _E+\frac{\left(I+f A V+\lambda _P P\right) \left(b_E+k_E E\right)}{K_B+\left(I+f A V+\lambda _P P\right)}-\xi\frac{  Q^n E}{q_c^n+Q^n}-d_E E \\ 
           & \qquad +\lambda_M M \left(I+f_M A V+\lambda _{MP} P\right) \\
    \dot{Q}&=\kappa \frac{I+\lambda _{QP} P}{\phi +I+\lambda _{QP} P}-d_Q Q \\
    \dot{M}&=\left(\mu _M E+b_M \frac{\left(I+f A V+\lambda _P P\right) E}{K_B+\left(I+f A V+\lambda _P P\right)}\right) \left(1-\frac{M}{M_{\max }}\right) \\
    A&=\frac{k_1 A_1+k_2 A_2}{K+A_1+A_2} \\
    A_1& = \frac{A_1^{dose}}{Vol_1} \sum_i^3 e^{-\eta_1(t-(\tau_i+\omega_1))}H(t-(\tau_i+\omega_1))\\
    A_2& = \frac{A_2^{dose}}{Vol_2} \sum_i^3 e^{-\eta_2(t-(\tau_i+\omega_2))}H(t-(\tau_i+\omega_2))\\
    T(0)&=\frac{b_T}{d_T} \\
    V(0)&=V_0
  \end{split}
\end{equation}

The primary interactions are between CD4+ (helper) T-cells ($T$), viral load ($V$), and infected T-cells.
Following \cite{Funk2001}, we divide the infected cells into four distinct types: actively infected cells ($I$), latently infected cells ($L$), persistently infected cells ($P$), and defectively infected cells ($D$).
T-cells are produced at rate $b_T$ and undergo natural degradation at the first-order rate $d_T$.
In the absence of infection, this leads to a steady state concentration $T = \frac{b_T}{d_T}$ which we use as a natural baseline, $T(0)$.
When viruses come into contact with T-cells, they produce each of the four infected cell types at respective rates $f_I, f_L, f_P, f_D$.
Virus is produced by both actively infected cells and persistently infected cells at rates $p_I, p_P$ and cleared at rate $c$.
Clearance is enhanced by the presence of bNAbs at a rate given by $A$.
Each infected cell type naturally degrades at rates $d_I, d_L, d_P, d_D$. 

Because they lead to active production of new virus, both actively infected and persistently infected cells are targeted by active effector cells ($E$) and killed at second-order rates $m_I, m_P$.
While they do not produce virus, latently infected cells can act as a long-lived pool that is converted into actively infected cells at the first-order rate $a$.
Finally, defectively infected cells simply detract from the overall infection.

The immune response is modeled by the activated effector cells and a phenomenological immune exhaustion variable ($Q$) \cite{Johnson2011}.
Effector response is activated and deactivated at intrinsic rates of $\lambda_E$ and $d_E$, respectively.
In the presence of actively or persistently infected cells, the effector response is stimulated with maximal zeroth and first-order rates $b_E$ and $k_E$.
The interaction of bNAb therapy with free virus can also contribute to effector response with a fraction $f$.
The Michaelis constant for effector stimulation is given by $K_B$.

After an extended period of activation, the immune response of the effector cells becomes exhausted, and the ability to fight infection is diminished.
This is modeled by a decay of $E$ with maximal rate $\xi$.
This decay saturates with $Q$ according to Michaelis constant $q_c$ and Hill coefficient $n$.
Exhaustion is generated by a saturable function of infected cells with maximum rate $\kappa$ and Michaelis constant $\phi$ and is balanced by first-order decay $d_Q$. 

There is no consensus on the best way to model immunological memory.
We followed the third of the three general approaches outlined in \cite{hutchison_modelling_2004}, the ``immune network model'' where transformed B-cells become long-lived memory cells that respond almost instantaneously to antigen and also activated versions of themselves to rapidly suppress a secondary infection.
Although complicated, this follows classic immunological theory with respect to memory cells while remaining agnostic about the complicated soup of cytokines involved in immunological responses rather than relying too heavily on the dynamics of one in particular, such as interleukin-2 as in \cite{mclean_modelling_1994}.

Following Desikan et al.~\cite{Desikan2020}, we incorporate the effect of bNAb therapy $A$ using a Hill function to model the exclusive binding of two drugs $A_1, A_2$ according to the derivation in \cite{Meyer2019}.
Individual concentrations of the two drugs are modeled by simple first-order clearance rates $\eta_1, \eta_2$, tuneable delays $\omega_1, \omega_2$ between injection and presentation, and effective compartmental volumes $Vol_1, Vol_2$. Dosages $A_1^{dose}, A_2^{dose}$ are injected at the three times $\tau_i$.  
\subsubsection{SHIV System Data}
\label{sec:SHIVmethods_system_data}

The time-course of HIV viral load and helper T-cell counts (CD4+ hereafter) were taken from experiments on 26 female macaques carried out by Nishimura et al.\cite{nishimura_early_2017} and available in the appendix to \cite{Desikan2020} for a total of 1,237 data points (848 viral loads, 389 CD4+ counts).
It should perhaps be noted that the procedure in this paper of using two bNAb's in combination had previously been applied in four previous studies, but generally starting far more than the 3 days post-infection used by Nishimura's group (see summary in \cite{nishimura_mice_2017}); all were far more successful than cART treatments in establishing long-term control of the infection.
For detailed ethics statements regarding non-human primate research, see the original paper \cite{nishimura_early_2017}.

We followed Nishimura's conclusions that the responses fell into three general categories: untreated, controllers, and non-controllers.
To avoid confusion between ``untreated controls'' and ``controllers'' we refer to the latter two groups as non-/responders throughout this paper.
Because the viral trajectory in the ten untreated macaques followed roughly the same pattern over roughly the same short time horizon, their data were aggregated in subsequent analysis.
Representatives were selected for the other two categories of treated macaques as having both a typical trajectory and a dense sampling regimen (DEWL as the responder, and DF06 as the nonresponder).
For the responder, we excluded data obtained following experimental anti-CD8 disruption approximately 2.5 years post-infection.
There is no reason to think similar results would not have been obtained using different representatives, though it may have been more difficult to reach convergence with fewer data points.
Following Desikan et al.\cite{Desikan2020}, we did not attempt to simultaneously fit CD4+ cell data. 

\subsubsection{SHIV Calibrated Model}
\label{sec:SHIVmethods_calibrated}

We calibrated our model to data from the experiments performed by Nishimura et al.\cite{nishimura_early_2017} as just described.
Data consists of plasma viral load measurements $V$, CD$4^+$ T-cell levels $T$, bNAb concentrations $A_1$, $A_2$ at observation times $t$. 
Additionally, for each time series we know the bNAb boluses $A_1^{dose}, A_2^{dose}$ and injection times $\tau_i$.
Because the T-cell data are noisy and inconsistent, we follow Desikan et al. \cite{Desikan2020} and do not include them when fitting.
Many viral load measurements lie below the detection limit of 100 RNA copies/mL; we also omit these data points.  

In many QSP models, including those we assimilated into our literature model, some or all parameter values are pinned to literature values.
While this is a common strategy for managing parametric complexity, we take the opposite approach in order to stress test our proposed methodology.
We obtained best-fit parameters values for each of the three times series of viral load: aggregated untreated, DEWL (responder), and DF06 (nonresponder).
We found two statistically equivalent nonresponder fits, DF06A and DF06B, that that use slightly different mechanisms and reproduce different qualitative features of the data.
DF06A's simplified drug dynamics capture the early rebound peak, but the long-term increase in V is not monotonic.
DF06B's renormalization of viral clearance leads to  smoother long-term growth but underestimates the early peak.
In retrospect, DF06B contributed little to further understanding of our model, but including multiple mechanisms is a useful strategy in other contexts, and we keep it here to demonstrate the methodology.

\subsubsection{SHIV Queries and Reduced Models}
\label{sec:shivmethods_queries}

The model queries included viral load $V$ and T-cell $T$ responses to zero, small, normal, and large bNAb therapy doses, corresponding to $A_1^{dose}, A_2^{dose} = 0$, $6.5\times 10^3$, $6.5\times 10^4$, and $6.5\times 10^5$.
We also include time courses for $A_1$ and $ A_2$.
These responses were predicted for times spanning $t=[0,1000]$ days.

Using these queries, MBAM was applied for each of the four parameter sets described above.
The reduced model for the aggregated untreated model are given in Eq.~\eqref{eq:SHIVaggregateduntreated}, for DEWL (responder) in Eq.~\eqref{eq:SHIVDEWL}, for DF06A (nonresponder) in Eq.~\eqref{eq:SHIFDF06A}, and for DF06B (nonresponder) in Eq.~\eqref{eq:SHIVDF06B}.

\begin{align}
  \label{eq:SHIVaggregateduntreated}
  \begin{split}
    \dot{T}&=b_T-d_T T-f_I T V \\
    V&=\frac{\widetilde{p_I} I}{\widetilde{c}+A} \\
    \dot{I}&=f_I T V-\widetilde{E} I \\
    \dot{\widetilde{E}}&=\frac{\widetilde{b_E} I}{K_B+I}-\xi \frac{\widetilde{Q}^n \widetilde{E}}{1+\widetilde{Q}^n}-d_E \widetilde{E} \\
    \dot{\widetilde{Q}}&=\widetilde{\kappa} I - d_Q \widetilde{Q}\\
    A&=\widetilde{k_2} A_2 \\
    T(0)&=\frac{b_T}{d_T} e^{-\frac{f_I}{\widetilde{c}}} \\
    I(0)&=\frac{b_T}{d_T} \left(1-e^{-\frac{f_I}{\widetilde{c}}}\right) \\
    \widetilde{p_I} = \frac{p_I}{V_0},\widetilde{c} = \frac{c}{V_0},\widetilde{k_2} = \frac{k_2}{K V_0},\widetilde{E} &= m_I E,\widetilde{b_E} = m_I b_E,\widetilde{Q} = \frac{Q}{q_c}, \widetilde{\kappa} = \frac{\kappa}{q_c}    
  \end{split}
\end{align}

\begin{align}
  \label{eq:SHIVDEWL}
  \begin{split}
    \dot{T}&=b_T-d_T T-f_I T V \\
    V&=\frac{\widetilde{p_I} I}{\widetilde{c}+A} \\
    \dot{I}&=f_I T V-d_I I-\widetilde{E} I \\
    \dot{\widetilde{E}}&=\frac{I \left(\widetilde{b_E}+k_E \widetilde{E}\right)}{K_B+I}-\xi \frac{\widetilde{Q}^n \widetilde{E}}{1+\widetilde{Q}^n}-d_E \widetilde{E} \\
    \dot{\widetilde{Q}}&=\widetilde{\kappa}  \frac{I}{\phi +I}-d_Q \widetilde{Q} \\
    A&=\widetilde{k_2} A_2 \\
    T(0)&=\frac{b_T}{d_T} e^{-\frac{f_I}{\widetilde{c}}} \\
    I(0)&=\frac{b_T}{d_T} \left(1-e^{-\frac{f_I}{\widetilde{c}}}\right) \\
    \widetilde{p_I} = \frac{p_I}{V_0},\widetilde{c} = \frac{c}{V_0},\widetilde{k_2} = \frac{k_2}{K V_0},\widetilde{E} &= m_I E,\widetilde{b_E} = m_I b_E,\widetilde{Q} = \frac{Q}{q_c}, \widetilde{\kappa} = \frac{\kappa}{q_c}
  \end{split}
\end{align}

\begin{align}
  \label{eq:SHIFDF06A}
  \begin{split}
    \dot{T}&=b_T-d_T T-\left(f_I+f_L\right) T V \\
    \dot{V}&=p_I I-(c+A) V \\
    \dot{I}&=f_I T V-d_I I+a L-\widetilde{E} I \\
    \dot{L}&=f_L T V-a L \\
    \dot{\widetilde{E}}&=\widetilde{\lambda_E}-d_E \widetilde{E} \\
    A&=\widetilde{k_2} A_2 \\
    T(0)&=\frac{b_T}{d_T} \\
    V(0)&=V_0 \\
    \widetilde{E} &= m_I E,\widetilde{\lambda_E} = m_I \lambda_E, \widetilde{k_2} = \frac{k_2}{K}
  \end{split}
\end{align}

\begin{align}
\label{eq:SHIVDF06B}
  \begin{split}
    \dot{T}&=b_T-d_T T-\left(f_I+f_L\right) T V \\
    V&=\frac{\widetilde{p_I} I}{\widetilde{c}+A} \\
    \dot{I}&=f_I T V-d_I I+a L-\widetilde{E} I \\
    \dot{L}&=f_L T V-a L \\
    \dot{\widetilde{E}}&=\widetilde{\lambda_E}-d_E \widetilde{E} \\
    A&=\frac{\widetilde{k_2} A_2}{K+A_1+A_2} \\
    T(0)&=\frac{b_T}{d_T} e^{-\frac{f_I+f_L}{\widetilde{c}}} \\
    I(0)&=\frac{f_I}{f_I+f_L}\frac{b_T}{d_T}\left(1-e^{-\frac{f_I+f_L}{\widetilde{c}}}\right) \\
    L(0)&=\frac{f_L}{f_I+f_L}\frac{b_T}{d_T}\left(1-e^{-\frac{f_I+f_L}{\widetilde{c}}}\right) \\
    \widetilde{p_I} = \frac{p_I}{V_0},\widetilde{c} = \frac{c}{V_0},\widetilde{k_2} &= \frac{k_2}{V_0},\widetilde{E} = m_I E,\widetilde{\lambda_E} = m_I \lambda_E
  \end{split}
\end{align}

We identify models of intermediate complexity, known as supremum, following the methods in of Petrie et al.\cite{petrie2022selecting}.
Conceptually, the supremum of two reduced models is found by applying the approximations that were common to both reductions.
The supremum of the four reduced models in Eqs.~\eqref{eq:SHIVaggregateduntreated}-~\eqref{eq:SHIVDF06B} is given in Eq.~\eqref{eq:SHIVSupremum}.
The supremum of the aggregated untreated and DEWL models is in Eq.~\eqref{eq:SHIVsupremum_untreated_DEWL} and for the two nonresponders (DF06A and DF06B) in Eq.~\eqref{eq:SHIVsupremumDF06}.

\begin{align}
  \label{eq:SHIVSupremum}
  \begin{split}
    \dot{T}&=b_T-d_T T-\left(f_I+f_L\right) T V \\
    \dot{V}&=p_I I-(c+A) V \\
    \dot{I}&=f_I T V-d_I I+a L-\widetilde{E} I \\
    \dot{L}&=f_L T V-a L \\
    \dot{\widetilde{E}}&=\widetilde{\lambda_E}+\frac{I \left(\widetilde{b_E}+k_E \widetilde{E}\right)}{K_B+I} - \xi\frac{\widetilde{Q}^n \widetilde{E}}{1+\widetilde{Q}^n}-d_E \widetilde{E} \\
    \dot{\widetilde{Q}}&=\widetilde{\kappa}  \frac{I}{\phi +I}-d_Q \widetilde{Q} \\
    A&=\frac{k_2 A_2}{K+A_1+A_2} \\
    T(0)&=\frac{b_T}{d_T} \\
    V(0)&=V_0 \\
    \widetilde{E} = m_I E,\widetilde{b_E} &= m_I b_E,\widetilde{\lambda_E} = m_I \lambda_E, \widetilde{Q} = \frac{Q}{q_c}, \widetilde{\kappa} = \frac{\kappa}{q_c}
  \end{split}
\end{align}

\begin{align}
  \label{eq:SHIVsupremum_untreated_DEWL}
  \begin{split}
    \dot{T}&=b_T-d_T T-f_I T V \\
    V&=\frac{\widetilde{p_I} I}{\widetilde{c}+A} \\
    \dot{I}&=f_I T V-d_I I-\widetilde{E} I \\
    \dot{\widetilde{E}}&=\frac{I \left(\widetilde{b_E}+k_E \widetilde{E}\right)}{K_B+I}-\xi \frac{\widetilde{Q}^n \widetilde{E}}{1+\widetilde{Q}^n}-d_E \widetilde{E} \\
    \dot{\widetilde{Q}}&=\widetilde{\kappa}  \frac{I}{\phi +I}-d_Q \widetilde{Q} \\
    A&=\widetilde{k_2} A_2 \\
    T(0)&=\frac{b_T}{d_T} e^{-\frac{f_I}{\widetilde{c}}} \\
    I(0)&=\frac{b_T}{d_T} \left(1-e^{-\frac{f_I}{\widetilde{c}}}\right), \\
    \widetilde{p_I} = \frac{p_I}{V_0},\widetilde{c} = \frac{c}{V_0},\widetilde{k_2} = \frac{k_2}{K V_0},\widetilde{E} &= m_I E,\widetilde{b_E} = m_I b_E,\widetilde{Q} = \frac{Q}{q_c}, \widetilde{\kappa} = \frac{\kappa}{q_c}
  \end{split}
\end{align}

\begin{align}
  \label{eq:SHIVsupremumDF06}
  \begin{split}
    \dot{T}&=b_T-d_T T-\left(f_I+f_L\right) T V \\
    \dot{V}&=p_I I-(c+A) V \\
    \dot{I}&=f_I T V-d_I I+a L-\widetilde{E} I \\
    \dot{L}&=f_L T V-a L \\
    \dot{\widetilde{E}}&=\widetilde{\lambda_E}-d_E \widetilde{E} \\
    A&=\frac{k_2 A_2}{K+A_1+A_2} \\
    T(0)&=\frac{b_T}{d_T} \\
    V(0)&=V_0 \\
    \widetilde{E} & = m_I E, \widetilde{\lambda_E} = m_I \lambda_E 
  \end{split}
\end{align}

Finally, we consider a subset set of these queries, and focus only on the profile of viral load for no dose and the normal dose.
The reduced model for aggregated untreated model for this simplified query is in Eq.~\eqref{eq:SHIVUTUQreduced}.

\begin{align}
  \label{eq:SHIVUTUQreduced}
  \begin{split}
    \dot{\widetilde{T}}&=1-d_T \widetilde{T}-f_I \widetilde{T} V \\
    V&=\frac{\widetilde{p_I} \widetilde{I}}{\widetilde{c}+A} \\
    \dot{\widetilde{I}}&=f_I \widetilde{T} V-\widetilde{E} \widetilde{I} \\
    \dot{\widetilde{E}}&=\frac{\widetilde{b_E} \widetilde{I}}{\widetilde{K_B}+\widetilde{I}}-\xi \frac{\widetilde{Q}^n \widetilde{E}}{1+\widetilde{Q}^n}-d_E \widetilde{E} \\
    \dot{\widetilde{Q}}&=\widetilde{\kappa} \widetilde{I}-d_Q \widetilde{Q}\\
    A&=\widetilde{k_2} A_2 \\
    \widetilde{T}(0)&=\frac{1}{d_T} e^{-\frac{f_I}{\widetilde{c}}} \\
    \widetilde{I}(0)&=\frac{1}{d_T} \left(1-e^{-\frac{f_I}{\widetilde{c}}}\right) \\
    \widetilde{p_I} = \frac{p_I b_T}{V_0},
    \widetilde{K_B} = \frac{K_B}{b_T},
    \widetilde{c} = \frac{c}{V_0},\widetilde{k_2} &= \frac{k_2}{K V_0},\widetilde{b_E} = m_I b_E,\widetilde{Q} = \frac{Q}{q_c}, \widetilde{\kappa} = \frac{\kappa b_T}{q_c}, \\
    \widetilde{T} = \frac{T}{b_T},\widetilde{I} = \frac{I}{b_T},\widetilde{E} &= m_I E,
  \end{split}
\end{align}

\begin{table}
    \centering
    \begin{tabular}{|c|c|c|c|}
         \hline
         Untreated & DEWL & DF06A & DF06B \\
         \hline
         $d_D \rightarrow 0$ & $d_D \rightarrow 0$ & $d_D\rightarrow 0$ & $\lambda_{QP}\rightarrow 0$ \\
         \hline
         $\lambda_{QP} \rightarrow 0$ & $\lambda_P \rightarrow 0$ & $\lambda_{QP}\rightarrow 0$ &  $d_D\rightarrow 0$\\
         \hline
         $p_P \rightarrow 0$ & $p_P \rightarrow 0$ & $\lambda_M\rightarrow 0$ &  $p_P\rightarrow 0$\\
         \hline
         $\lambda_M \rightarrow 1$ & $f_D \rightarrow 0$ & $f_M\rightarrow 0$ &  $\lambda_M\rightarrow 0$\\
         \hline
         $f_M \rightarrow 0$ & $\lambda_M \rightarrow 0$ & $b_M\rightarrow 0$ &  $f_M\rightarrow 0$\\
         \hline
         $b_M \rightarrow 0$ & $f_M \rightarrow 0$ & $\lambda_{MP}\rightarrow 0$ &  $b_M\rightarrow 0$\\
         \hline
         $\lambda_{MP} \rightarrow 0$ & $b_M \rightarrow 0$ & $\xi\rightarrow 0$ &  $\lambda{MP}\rightarrow 0$\\
         \hline
         $\lambda_P \rightarrow 1$ & $\lambda_{MP}\rightarrow 0$ & $M_\textrm{max}\rightarrow \infty$ &  $M_\textrm{max}\rightarrow \infty$\\
         \hline
         $f_D \rightarrow 1$ & $f_P \rightarrow 1$ & $q_c\rightarrow 1$ &  $\mu_M\rightarrow 0$\\
         \hline
         $M_\textrm{max} \rightarrow \infty$ & $\mu_M \rightarrow 0$ & $\mu_M\rightarrow 0$ & $\xi\rightarrow 0$ \\
         \hline
         $\mu_M \rightarrow 0$ & $M_\textrm{max} \rightarrow \infty$ & $d_Q\rightarrow 0$ &  $\lambda_P\rightarrow 0$\\
         \hline
         $q_c \rightarrow 1$ & $d_L \rightarrow 0$ & $\kappa\rightarrow 0$ &  $q_c,\kappa\rightarrow \infty$\\
         \hline
         $V_0, p_I, c, f, k_1, k_2 \rightarrow \infty$ & $d_P \rightarrow 0$ & $\phi\rightarrow \infty$ &  $\widetilde{\kappa}\rightarrow 0$\\
         \hline
         $m_P \rightarrow 0$ & $m_I \rightarrow 1$ & $f_D \rightarrow 0$ &  $d_Q\rightarrow 0$\\
         \hline
         $m_I \rightarrow 1$ & $\kappa, q_c \rightarrow \infty$ & $d_L\rightarrow 0$ &  $d_P\rightarrow 0$\\
         \hline
         $d_I \rightarrow 0$ & $m_P \rightarrow 0$ & $d_P\rightarrow 0$ &  $m_P\rightarrow 0$\\
         \hline
         $d_L \rightarrow 0$ & $\widetilde{f_P} \rightarrow 0$ & $b_E\rightarrow 0$ &  $\phi\rightarrow 0$\\
         \hline
         $d_P \rightarrow 0$ & $\lambda_E \rightarrow 0$ & $m_P \rightarrow 0$ &  $f_D\rightarrow 0$\\
         \hline
         $\lambda_E \rightarrow 0$ & $f\rightarrow 0$ & $p_P\rightarrow 0$ &  $d_L\rightarrow 0$\\
         \hline
         $f_P \rightarrow 0$ & $V_0,p_I,c,k_1,k_2\rightarrow \infty$ & $f\rightarrow 0$ &  $b_E\rightarrow 0$\\
         \hline
         $\widetilde{k_1} \rightarrow 0$ & $\widetilde{k_1}\rightarrow 0$ & $k_E\rightarrow 0$ &  $k_E\rightarrow 0$\\
         \hline
         $\kappa, \phi \rightarrow \infty$ & $f_L \rightarrow 0$ & $\lambda_P\rightarrow 0$ &  $k_1\rightarrow 0$\\
         \hline
         $\widetilde{k_2}, K \rightarrow \infty$ & $a\rightarrow 0$ & $f_P\rightarrow 0$ &  $f\rightarrow 0$\\
         \hline
         $\omega_1 \rightarrow 0$ & $\widetilde{k_2},K \rightarrow \infty$ & $KB\rightarrow \infty$ &  $KB\rightarrow 0$\\
         \hline
         $k_E \rightarrow 0$ &  & $k_1\rightarrow 0$ &  $f_P\rightarrow 0$\\
         \hline
         $\widetilde{f} \rightarrow 0$ & & $m_I\rightarrow 1$ &  $m_I\rightarrow 1$\\
         \hline
         $f_L \rightarrow 0$ &  & $k_2, K\rightarrow \infty$ &  $V_0,p_I,c,k_2\rightarrow \infty$\\
         \hline
         $a \rightarrow 0$ &  &  &  \\
         \hline
    \end{tabular}
    \caption{\textbf{MBAM Reductions} for each of the four calibrated SHIV models.
    Parameters set to 1 are structurally unidentifiable and are pinned to an arbitrary value.}
    \label{tab:reductions}
\end{table}

\begin{table}[ht]
\centering
\begin{tabular}{|c|c|c|c|}
\hline
\multicolumn{4}{|c|}{\textbf{Supremum Reductions}} \\
\hline
\multicolumn{4}{|c|}{$d_D \rightarrow 0$} \\
\multicolumn{4}{|c|}{$f_D \rightarrow 0$} \\
\multicolumn{4}{|c|}{$\lambda_M \rightarrow 0$} \\
\multicolumn{4}{|c|}{$f_M \rightarrow 0$} \\
\multicolumn{4}{|c|}{$b_M \rightarrow 0$} \\
\multicolumn{4}{|c|}{$\lambda_{MP} \rightarrow 0$} \\
\multicolumn{4}{|c|}{$\mu_M \rightarrow 0$} \\
\multicolumn{4}{|c|}{$M_{\max} \rightarrow \infty$} \\
\multicolumn{4}{|c|}{$\lambda_P \rightarrow 0$} \\
\multicolumn{4}{|c|}{$p_P \rightarrow 0$} \\
\multicolumn{4}{|c|}{$d_P \rightarrow 0$} \\
\multicolumn{4}{|c|}{$m_P \rightarrow 0$} \\
\multicolumn{4}{|c|}{$f_P \rightarrow 0,\ \lambda_{QP} \rightarrow \infty$} \\
\multicolumn{4}{|c|}{$\widetilde{f_P} \rightarrow 0$} \\
\multicolumn{4}{|c|}{$m_I,\ m_P \rightarrow \infty,\ \lambda_E,\ b_E \rightarrow 0$} \\
\multicolumn{4}{|c|}{$\kappa \rightarrow \infty,\ q_c \rightarrow \infty$} \\
\multicolumn{4}{|c|}{$d_L \rightarrow 0$} \\
\multicolumn{4}{|c|}{$f \rightarrow 0$} \\
\multicolumn{4}{|c|}{$k_1 \rightarrow 0$} \\
\hline
Untreated & Responder & Nonresponder A & B \\
\hline 
\multicolumn{2}{|c|}{$\lambda_E \rightarrow 0$} & \multicolumn{2}{c|}{$\xi \rightarrow 0$} \\
\multicolumn{2}{|c|}{$V_0, p_I, c, k_1, k_2 \rightarrow \infty$} & \multicolumn{2}{c|}{$\kappa \rightarrow 0$} \\
\multicolumn{2}{|c|}{$f_L \rightarrow 0$} & \multicolumn{2}{c|}{$d_Q \rightarrow 0$} \\
\multicolumn{2}{|c|}{$a \rightarrow 0$} & \multicolumn{2}{c|}{$\phi \rightarrow 0$} \\
\multicolumn{2}{|c|}{$k_2, K \rightarrow \infty$} & \multicolumn{2}{c|}{$b_E \rightarrow 0$} \\
\cline{1-2}
$d_I \rightarrow 0$ & & \multicolumn{2}{c|}{$k_E \rightarrow 0 $} \\
\cline{3-4}
$\kappa, \phi  \rightarrow \infty$ & & $k_2, K \rightarrow \infty$ & $V_0, p_I, c, k_2 \rightarrow \infty$ \\
$\omega_1 \rightarrow 0$ & & & \\
$k_E \rightarrow 0$ & & & \\
\hline
\end{tabular}
\caption{\textbf{Supremum reordered MBAM reductions.}  
Collecting approximations common to different reductions leads to their ``supremum models."
}
\end{table}

\subsubsection{SHIV Target Identification}
\label{sec:SHIVmethods_targetid}

QSP model simulations are well-suited for drug discovery due to the relative ease with which a biological system's response to perturbations can be measured in detail.
Ideally, specific mechanistic targets can be identified that play a significant role in disease progression.
We demonstrate this task in the SHIV model by locating system parameters for which small changes lead to a differentiated predicted response.
Specifically, we focus on the nonresponder DF06A for which bNAb therapy on its own is ineffective.

Calculations are performed using the supremum model in Eq.~\eqref{eq:SHIVSupremum} as it is the simplest possible model that exhibits both responder and nonresponder behavior.
We formulate an optimization problem to determine the minimal parameter perturbations leading to a change of response:
\begin{align}
   \min_\theta \sum_i\sinh^{-1}\left( V_i(\theta)\right)^2 + \lambda\sum_\mu |\theta_\mu - \theta_\mu^{\textrm{nonresp}}|
\end{align}
where the sum over $i$ corresponds to different time points at which the viral load $V_i$ is predicted, the sum over $\mu$ corresponds to the different parameters $\theta_\mu$, and $\lambda$ is a Lagrange multiplier.
Conceptually, we seek parameters minimally perturbed from those of the nonresponder that lead to a small viral load.
In the first term, we used an inverse $\sinh$ function that behaves like a logarithm for large argument and nearly linear for small argument.
In this way, it acts to compress the large range of $V$ while (unlike the log function) retaining a finite lower bound of zero. 
We use an $\ell_1$ norm in the second term to encourage sparsity in the solution: we want to target as few parameters as possible.
Solving this optimization problem gives a sparse set of parameters which are the most influential on treatment outcome.
We then conduct subsequent iterations to search for less intuitive targets that may be easier to realize biochemically.
To search for other targets, we hold the initially identified parameters fixed at their nominal values and repeat the optimization over the remaining parameters. 
Iteratively holding some parameters fixed results in a number of different parameter combination targets.

While the optimization targets required perturbing a few parameters, we found that marginally larger perturbations of individual parameters were also sufficient to induce the desired response.

\subsubsection{SHIV virtual population generation}
\label{sec:SHIVmethods_uq}

QSP models are frequently used to make extrapolative predictions beyond the context of the data from which they are calibrated.
For example, a patient model can be queried to investigate the response and side effects of a potential therapeutic intervention.
While a parameterized model will always output some prediction, it is crucial to quantify the confidence associated with it in order to properly assess its reliability.
While there are many sources of uncertainty in mathematical modeling~\cite{zhang_basic_2020}, QSP models are often dominated by parametric uncertainty~\cite{sharp_parameter_2022}.
Here, we propagate forward the parameter uncertainty resulting from the fitting process to downstream predictions of interest by generating a virtual population. 

We generate our virtual population by sampling the Bayesian posterior distribution to propagate uncertainty directly to the predictions. The posterior is the conditional probability distribution of the parameters given the observed data, and is calculated using Bayes' rule: 
\begin{equation}
    \label{eq:bayes}
    \rho( \theta \vert x ) \propto \rho( x \vert \theta ) \rho(\theta).
\end{equation}
The likelihood $\rho(x\vert\theta)$ encodes the noise model, which we take as i.i.d. normal, while the prior $\rho(\theta)$ weights parameter values in accordance with existing knowledge. 

To sample the Bayesian posterior $\rho(\theta\vert x)$, we use the affine invariant Markov chain Monte Carlo (MCMC) method \cite{AIMCMC}.
We use an uninformative flat prior centered on the parameter best-fit values.
To ensure that the posterior is proper, this flat prior is bounded in each log-parameter at $\pm8$ about the center. 
Our Markov chain used 120 walkers and ran for 10000 steps after an initial burn-in of 1000 steps. We sampled this chain every 100 steps to eliminate correlations.
After the posterior was sampled, we calculated the corresponding model predictions for each member of this virtual population. 

We demonstrate our workflow for virtual population generation by predicting the effect of bNAb therapy on the patients for which only untreated data is available.
Using the model in Eq.~\ref{eq:SHIVUTUQreduced}, we fit the model to the available data and generate a virtual population using the Bayesian method. 
We then categorize the predictions for each virtual patient according to clinical outcome. Differential HIV infection outcomes have been hypothesized to arise from dynamical bistability \cite{Baral2019,Desikan2020}. Consistent with this established characterization, we determine the steady state stability of each virtual patient.

For our reduced model, we find that dynamics are exclusively monostable or bistable. We therefore first divide the virtual population by this categorization. Because they have only a single steady state, monostable patients cannot exhibit a long-term response to treatment. Bistable patients, however, can end up in either a high or low viral load steady state. Figure~\ref{fig7}a compares the dynamics of a typical monostable patient to a bistable responder. The monostable dynamics show a transient response to treatment, while the responder dynamics demonstrate that treatment has perturbed the system from the unhealthy high steady state to the controlled lower steady state.

Bistable virtual patients can be further divided according to their clinical outcomes both with and without treatment, as shown in Figure~\ref{fig7}b-f. Immune patients approach the lower steady state both with and without treatment. Untreated responders reach the high steady state, but switch to the low steady state with treatment. Nonresponders approach the high steady state regardless of treatment. Finally, untreated adverse patients approach the healthy low steady state, but are perturbed by treatment to reach the unhealthy high steady state.

Stability is determined by numerically calculating the system fixed points and counting the number of positive solutions---2 for monostable patients, and 4 for bistable patients. Bistable patient responses are categorized by comparing the viral load at day 1000 to the two stable fixed points and determining which is closer.

\section{Acknowledgments}
We thank J.~Bai and S.~Nayak for suggesting we consider coagulation, and their expertise in creating the LDM and interpreting its reductions.
Our thanks to H.~Moore for her wide knowledge of QSP modeling, and invaluable suggestions about the ways we should distinguish our workflow from others. 

\bibliographystyle{plos2015}



\bibliography{SHIV_refs}

\end{document}